\documentclass[referee]{raa}            

\usepackage{graphicx,times}             
\usepackage[round]{natbib}
\usepackage{amssymb}
\usepackage{txfonts}
\usepackage{graphicx}
\usepackage{multirow}
\usepackage{dcolumn}
\usepackage{array}
\usepackage{xcolor}
\usepackage{lscape}

\begin{document}
\title{Surface thermophysical properties investigation of the potentially hazardous asteroid (99942) Apophis}

\volnopage{Vol.0 (200x) No.0, 000--000}      
\setcounter{page}{1}          

\author{LiangLiang Yu\inst{1} \and Jianghui Ji\inst{2} \and Wing-Huen Ip\inst{1,3}}

\institute{
Space Science Institute, Macau University of Science and Technology, Taipa, Macau;
{\it yullmoon@live.com}\\
\and
CAS Key Laboratory of Planetary Sciences, Purple Mountain Observatory, Chinese Academy of Sciences, Nanjing 210008, China;
{\it jijh@pmo.ac.cn} \\
\and
Institute of Astronomy, National Central University, Jhongli, Taoyuan City 32001, Taiwan;
{\it wingip@astro.ncu.edu.tw}\\
}

\date{Received~~2009 month day; accepted~~2009~~month day}

\abstract{
In this work, we investigate the surface thermophysical properties (thermal
emissivity, thermal inertia, roughness fraction and geometric albedo) of
asteroid (99942) Apophis, using the currently available thermal infrared
observations of CanariCam on Gran Telescopio CANARIAS and far-infrared
data by PACS of Herschel, on the basis of the Advanced thermophysical model.
We show that the thermal emissivity of Apophis should be wavelength dependent
from $8.70~\mu m$ to $160~\mu m$, and the maximum emissivity may arise around
$20~\mu m$ similar to that of Vesta. Moreover, we further derive the thermal
inertia, roughness fraction, geometric albedo and effective diameter of Apophis
within a possible 1$\sigma$ scale of
$\Gamma=100^{+100}_{-52}\rm~Jm^{-2}s^{-0.5}K^{-1}$, $f_{\rm r}=0.78\sim1.0$,
$p_{\rm v}=0.286^{+0.030}_{-0.026}$, $D_{\rm eff}=378^{+19}_{-25}\rm~m$,
and 3$\sigma$ scale of $\Gamma=100^{+240}_{-100}\rm~Jm^{-2}s^{-0.5}K^{-1}$,
$f_{\rm r}=0.2\sim1.0$, $p_{\rm v}=0.286^{+0.039}_{-0.029}$,
$D_{\rm eff}=378^{+27}_{-29}\rm~m$. The derived low thermal inertia but high
roughness fraction may imply that Apophis could have regolith on its surface,
and less regolith migration process has happened in comparison with asteroid
Itokawa. Our results show that small-size asteroids could also have fine
regolith on the surface, and further infer that Apophis may be delivered
from the Main Belt by Yarkovsky effect.
}

\authorrunning{L.L. Yu, J.H. Ji \& W. H. Ip }            
\titlerunning{Surface thermophysical properties of Apophis}  

\maketitle
\section{Introduction}
(99942) Apophis (2004 MN4) is categorized as an Aten-group near-Earth
asteroid (NEA) based on its orbital characteristics. The asteroid was known
as a potentially hazardous object (PHO) with a significant high Earth impact
probability of  $~$2\% in 2029, based on the observations of December, 2004
reported by JPL-Sentry and ESA-NEODyS. On December 27, 2004,
the Spacewatch survey announced pre-discovery observations \citep{Gilmore2004}
from March 2004, and removed any impact possibility in 2029.

Radar observations of Apophis were obtained by the Arecibo observatory
in January 2005, August 2005 and May 2006. More recently, \citet{Thuillot2015}
reported new astrometric observations obtained by space-based $Gaia$-$FUN$-$SSO$
during the Apophis' latest period of visibility from December 21, 2012 to
February 5, 2013, and soon afterwards, \citet{Wang2015} published precise 298
CCD position data observed by the ground-based 2.4-m telescope at Yunnan Observatory
from February 4, 2013 to March 2, 2013. These observations significantly reduced
Apophis' orbital uncertainty and led to a more accurate estimation of the
encountering distance about 5-6 Earth radius from the geocenter in 2029
\citep{Gilmore2008,Thuillot2015,Wang2015}. Although the 2029 impact has been
eliminated, other potential impacts may still exist in the following decades.

\citet{Chesley2005} showed that the Yarkovsky effect \citep{Bottke2006}
significantly affects post-2029 predictions of Apophis' orbit evolution
and thus should be taken into account for its impact predictions. As well
known, the Yarkovsky effect due to the recoil of reflected and thermal
emitted photons is one most important of those non-gravitational effects.
Utilizing selected best astrometric and radar data covering the interval
from  March 15, 2004 to December 29, 2012, \citet{Farnocchia2013} carried out a
detail orbital analysis and quantified that predictions of the Earth impacts
of Apophis between 2060 and 2105 are sensitive to its physical parameters,
including diameter, albedo, rotation period, obliquity, density, and thermal
inertia, which determine the rate of Yarkovsky drift of Apophis' semimajor axis.
In addition, \citet{Vokrouhlicky2015} provided a more advanced estimation of
Apophis' Yarkovsky effect, and predicted that Apophis' secular change in the
semimajor axis may be $(-12.8\pm3.6)\times10^{-4}\rm~AU/Myr$. Therefore accurate
thermal properties (thermal inertia, roughness fraction and so on) and shape
model are necessary to predict its Earth impact probability in the following
decades in consideration of the significantly important Yarkovsky effect.

By comparing spectral and mineralogical characteristics of likely meteorite analogs
from 0.55 to 2.45 $\mu$m reflectance spectral measurements of Apophis observed by NASA
IRTF and Baade Telescope at the Magellan Observatory, \citet{Binzel2009} found that
Apophis appears well classified as an Sq-type and most closely resembles LL ordinary
chondrite meteorites, which is rather similar to asteroid Itokawa \citep{Abe2006}.

With the photometric observations of Apophis from December 2012 to April 2013,
\citet{Pravec2014} showed Apophis has a non-principle axis rotation, and
rebuilt its convex shape model, where the retrograde rotation period is
$P_{1}=30.56~$h and the spin axis orientation is ($-75.0^{\circ}$, $250.0^\circ$).
\citet{Pravec2014} also derived Apophis' average absolute visual magnitude to be
$H_{\rm v}=19.09$ assuming the slope parameter to be $G=0.24\pm0.11$.

Generally the thermal inertia of an asteroid can be evaluated by fitting
thermal infrared observations with a thermophysical model to reproduce
thermal emission curves. Following the shape model of \citet{Pravec2014},
\citet{Muller2014} utilized the so-called thermophysical model (TPM) to
analyze Apophis' Herschel-far-infrared observations, and derived its
thermal inertia to be $250-800\rm~Jm^{-2}s^{-0.5}K^{-1}$ with a best
solution about $600\rm~Jm^{-2}s^{-0.5}K^{-1}$, mean effective diameter
$375^{+14}_{-10}$ m, and geometry albedo $0.30^{+0.05}_{-0.06}$.
All these properties of Apophis are very similar to those of Itokawa
\citep{Muller2005,Fujiwara2006}, indicating that Apophis might also have
a rubble-pile structure. However, recently \citet{Licandro2016} derived a
relatively lower thermal inertia $50-500\rm~Jm^{-2}s^{-0.5}K^{-1}$ with
a best solution of $\Gamma=150\rm~Jm^{-2}s^{-0.5}K^{-1}$ and low surface
roughness, indicating a surface material like that of Eros. \citet{Licandro2016}
used the same TPM like \citet{Muller2014}, but added three new mid-infrared
data measured by CanariCam of Gran Telescopio CANARIAS. Both the work of
\citet{Muller2014} and \citet{Licandro2016} assumed constant thermal
emissivity to derive the surface thermophysical properties, but actually
the thermal emissivity may be wavelength dependent from mid-infrared to
far infrared \citep{Muller1998}, which should be taken into account when
deriving the surface thermophysical properties.

In the present work, we utilize the independently developed thermophysical
simulation codes \citep{Yu2014, Yu2015} based on the frame work of Advanced
Thermophysical Model (ATPM) \citep{Rozitis2011} to analyse the mid-infrared
and far infrared data of Apophis observed by CanariCam and Herschel respectively.
We show that if wavelength dependent emissivities are used, better solutions
can be obtained, and the derived thermal inertia and effective diameter can be
different from the results derived on the basis of constant emissivity. Moreover,
we find that the surface of Apophis may be a high rough surface with low thermal
inertia materials. The low thermal inertia of the small-size asteroid Apophis
reveals that small-size asteroid may also have regolith on the surface, and further suggests that Apophis may be delivered from
the Main Belt by Yarkovsky effect.

\section{Thermophysical Modelling}
\subsection{Thermal infrared Observations}
Until by now, the available thermal infrared measurements of Apophis are the
far-IR data observed by the Herschel Space Observatory \citep{Muller2014}, and three
mid-IR data measured by CanariCam of Gran Telescopio CANARIAS \citep{Licandro2016}.
All these data are used in this work to be compared with the theoretical flux simulated
from the so-called Advanced thermophysical Model so as to derive the possible scale of
surface thermophysical properties. We tabulate data used in the fitting in Table \ref{obs}.

\begin{table*}[htbp]
\centering
\renewcommand\arraystretch{1.1}
\caption{Observational data used in this work. \citep{Muller2014,Licandro2016} }
\label{obs}
\begin{tabular}{@{}lcccccc@{}}
\hline
 UT & Wavelength & Flux  & $r_{\rm helio}$ & $\Delta_{\rm obs}$ & $\alpha$    & Observation \\
    & ($\mu m$) & (mJy) & (AU)            & (AU)              & ($^{\circ}$)  & Instrument \\
\hline
 2013-01-06 00:10 &   70.0 &  36.3$\pm$1.1 & 1.03593  & 0.096247 & +60.44 & Herschel/PACS\\
 2013-01-06 00:10 &  160.0 &   8.7$\pm$3.3 & 1.03593  & 0.096247 & +60.44 & Herschel/PACS\\
 2013-01-06 00:44 &  100.0 &  22.8$\pm$1.7 & 1.03599  & 0.096234 & +60.40 & Herschel/PACS\\
 2013-01-06 00:44 &  160.0 &   7.4$\pm$3.8 & 1.03599  & 0.096234 & +60.40 & Herschel/PACS\\
 2013-01-06 01:15 &   70.0 &  37.5$\pm$1.3 & 1.03604  & 0.096221 & +60.36 & Herschel/PACS\\
 2013-01-06 01:15 &  160.0 &   9.8$\pm$2.5 & 1.03604  & 0.096221 & +60.36 & Herschel/PACS\\
 2013-01-06 01:47 &  100.0 &  25.0$\pm$1.5 & 1.03609  & 0.096208 & +60.32 & Herschel/PACS\\
 2013-01-06 01:47 &  160.0 &   8.2$\pm$2.2 & 1.03609  & 0.096208 & +60.32 & Herschel/PACS\\
\\
 2013-03-14 06:40 &   70.0 &  12.6$\pm$2.7 & 1.093010 & 0.232276 & -61.38 & Herschel/PACS\\
 2013-03-14 06:54 &   70.0 &  11.4$\pm$2.7 & 1.093003 & 0.232307 & -61.38 & Herschel/PACS\\
 2013-03-14 07:07 &   70.0 &  10.4$\pm$2.7 & 1.092996 & 0.232338 & -61.39 & Herschel/PACS\\
 2013-03-14 07:21 &   70.0 &  12.5$\pm$2.6 & 1.092989 & 0.232368 & -61.39 & Herschel/PACS\\
 2013-03-14 07:35 &   70.0 &  13.3$\pm$2.7 & 1.092983 & 0.232397 & -61.40 & Herschel/PACS\\
 2013-03-14 07:49 &   70.0 &  12.4$\pm$2.6 & 1.092976 & 0.232427 & -61.40 & Herschel/PACS\\
\\
 2013-01-29 22:04 &   12.5 & 240.0$\pm$20  & 1.079706 & 0.113407 & -31.73 & CanariCam/GTC\\
 2013-01-29 23:09 &   17.65& 310.0$\pm$70  & 1.079761 & 0.113478 & -31.74 & CanariCam/GTC\\
 2013-01-29 23:52 &    8.70& 140.0$\pm$10  & 1.079816 & 0.113549 & -31.75 & CanariCam/GTC\\
\hline
\end{tabular}
\end{table*}

\subsection{Advanced thermophysical model}
In ATPM Procedure \citep{Rozitis2011, Yu2014, Yu2015}, we treat an asteroid to
be a polyhedron composed of $N$ triangle facets. For each facet, the roughness is
modelled by a fractional coverage of hemispherical micro-craters, symbolized by
$f_{\rm r}$ ($0\leq f_{\rm r}\leq 1$), whereas the remaining fraction, $1-f_{\rm r}$,
represents a smooth flat surface. For such rough surface facet, the conservation of
energy leads to an instant heat balance between sunlight, thermal emission, heat
conduction, multiple-scattered sunlight and thermal-radiated fluxes from other facets.
If each facet is small enough and far larger than the spatial scale of roughness,
the heat conduction in that region can be approximatively described as one-dimensional
(1D) heat conduction. Meanwhile, the temperature $T_{i}$ of each facet varies
with time as the asteroid rotates. In this process, $T_{i}$ can be significantly
affected by shading, multiple-scattered sunlight and thermal-radiated fluxes
from other facets, which well explains the so-called thermal infrared beaming
effect. When the entire asteroid comes into the final thermal equilibrium state,
$T_{i}$ will change periodically following the rotation of the asteroid. Therefore,
we can build numerical codes to simulate $T_{i}$ at any rotation phase for the
asteroid. For a given observation epoch, ATPM can reproduce a theoretical profile
to each observation flux as:
\begin{equation}
F_{\rm model}(\lambda)=\sum^{N}_{i=1}\epsilon(\lambda) S(i)f(i)B(\lambda, T_{i})~,
\label{fmodel}
\end{equation}
where $\epsilon(\lambda)$ is the monochromatic emissivity at wavelength $\lambda$,
$S(i)$ is the area of facet $i$, $f(i)$ is the view factor of facet $i$ to the telescope
\begin{equation}
f(i)=v_i\frac{\vec{n}_{i}\cdot\vec{n}_{\rm obs}}{\pi\Delta^{2}},~
\end{equation}
$v_{\rm i}=1$ indicates facet $i$ is visible (otherwise $v_{\rm i}=0$),
and $B(\lambda, T_{i})$ is the Planck function:
\begin{equation}
B(\lambda, T_{i})=\frac{2\pi hc^{2}}{\lambda^{5}}\frac{1}{ \exp\big(\frac{hc}{\lambda kT_{i}}\big)-1 }~.
\end{equation}
Thus the calculated $F_{\rm model}$ can be compared with the thermal
infrared fluxes summarized in Table \ref{obs} in the fitting process.

\subsection{Fitting Procedure}
In order to derive the thermophysical properties of Apophis via the ATPM
procedure, several physical parameters are needed, including the 3D shape
model, effective diameter $D_{\rm eff}$, and the so-called thermal parameter
\begin{equation}
\Phi=\frac{\Gamma\sqrt{\omega}}{\varepsilon\sigma T^{3}_{\rm eff}}~,
\end{equation}
where $\omega$ is the rotation frequency, $\Gamma$ is the thermal inertia,
$\varepsilon$ is the averaged thermal emissivity over the entire emission spectrum, and
\[T_{\rm eff}=\left[\frac{(1-A_{\rm B})F_{\odot}}{\varepsilon\sigma d_{\rm \odot}^2}\right]^{1/4}~,\]
is the effective temperature. The rotation frequency $\omega$ can be easily determined
from light curves, while thermal inertia $\Gamma$ is the parameter of interest which
would be treated as free parameter in the fitting procedure.

\begin{figure}[htbp]
\includegraphics[scale=0.6]{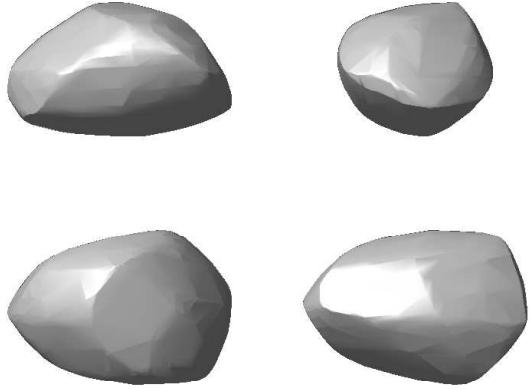}
  \centering
  \caption{The 3D shape model of Apophis utilized in this work is the light-curve inversion shape
  rebuilt by \citet{Pravec2014}.
  }\label{shape}
\end{figure}

We can employ the light-curve inversion 3D shape model of Apophis rebuilt
by \citet{Pravec2014} (Figure \ref{shape}) in our fitting procedure.
According to \citet{Fowler1992}, an asteroid's effective diameter $D_{\rm eff}$,
defined by the diameter of a sphere with a identical volume to that of the shape
model, can be related to its geometric albedo $p_{v}$ and absolute visual
magnitude $H_{v}$ via:
\begin{equation}
D_{\rm eff}=\frac{1329\times 10^{-H_{v}/5}}{\sqrt{p_{v}}}~(\rm km) ~.
\label{Deff}
\end{equation}
In addition, the geometric albedo $p_{v}$ is related to the effective
Bond albedo $A_{\rm eff,B}$ by
\begin{equation}
A_{\rm eff,B}=p_{v}q_{\rm ph}~,
\label{aeffpv}
\end{equation}
where $q_{\rm ph}$ is the phase integral that can be approximated by
\citep{Bowell}
\begin{equation}
q_{\rm ph}=0.290+0.684G~,
\label{qph}
\end{equation}
in which $G$ is the slope parameter in the $H, G$ magnitude system of
\citet{Bowell}. We use $H_{\rm v}=19.09\pm0.19$, $G=0.24\pm0.11$, the
results of \citet{Pravec2014} in the fitting procedure.

On the other hand, the asteroid's effective Bond albedo is the averaged
result of both the albedo of smooth and rough surface, which can be expressed
as the following relationship:
\begin{equation}
A_{\rm eff,B}=(1-f_{\rm r})A_{B}+f_{\rm r}\frac{A_{B}}{2-A_{B}}~,
\label{abfr}
\end{equation}
where $A_{B}$ is the Bond albedo of smooth lambertian surface, which should
be related to the composition of surface materials. Thus a input roughness
fraction $f_{\rm r}$ and geometric albedo $p_{\rm v}$ can lead to a unique
Bond albedo $A_{B}$ and effective diameter $D_{\rm eff}$ to be used to fit
the observations.

Then we have three free parameters --- thermal inertia, roughness fraction, and
geometric albedo (or effective diameter) that would be extensively investigated
in the fitting process. Other parameters are listed in Table \ref{phpa}.
\begin{table}[htbp]
 \centering
 \renewcommand\arraystretch{1.3}
 \caption{Assumed physical parameters used in ATPM.}
 \label{phpa}
 \begin{tabular}{@{}lcc@{}}
 \hline
 Property & Value & References \\
 \hline
 Number of vertices  &     1014               & \citep{Pravec2014}  \\
 Number of facets    &     2024               & \citep{Pravec2014}   \\
 Shape (a:b:c)       & 1.1135:1.0534:1        & \citep{Pravec2014}   \\
Angular momentum vector & ($-75.0^{\circ}$, $250.0^\circ$) & \citep{Pravec2014}  \\
 Spin period         &     30.56  h           & \citep{Pravec2014}   \\
 Absolute magnitude  &  $19.09\pm0.19$        & \citep{Pravec2014} \\
 Slope parameter     &  $0.24\pm0.11$         & \citep{Pravec2014} \\
 Thermal emissivity  &      0.9               &  assumption        \\
 \hline
\end{tabular}
\end{table}

It should be also noticed here that the utilized observation data are
observed at various wavelength from mid-infrared to far-infrared, thus
it may be no longer suitable to assume an average constant emissivity
for all wavelength when calculating flux by Equation (\ref{fmodel}),
because the spectral emissivity may differ with wavelength from mid-infrared
to far-infrared \citep{Muller1998}. Therefore the spectral emissivity $\epsilon$
remains to be a free parameter as well.

For the input free parameters, we use an initial geometric albedo $p_{\rm v}=0.3$,
but scan thermal inertia $\Gamma$ in the range $0\sim500\rm~Jm^{-2}s^{-0.5}K^{-1}$,
and roughness fraction $f_{\rm r}$ in the range $0.0\sim1.0$.
For the specral emissivity $\epsilon$, we firstly assume it to be constant 0.9
(CE for short), then vary it to see if we can get better solutions.
For each pair of ($\Gamma$, $f_{\rm r}$), we could find a new $p_{\rm v}$
(or $D_{\rm eff}$) that gives a minimum reduced $\chi^{2}$ defined as
\begin{equation}
\chi^{2}_{\rm r}=\frac{1}{n-3}\sum^{n}_{i=1}
\Big[\frac{F_{\rm model}(\lambda_i,\Gamma,f_{\rm r},p_{\rm v})
    -F_{\rm obs}(\lambda_{i})}{\sigma_{\lambda_{i}}}\Big]^{2}~,
\label{l2}
\end{equation}
which is used to assess the fitting degree of our model with respect to the observations.
Herein the predicted model flux $F_{\rm model}$ is a rotationally averaged profile
assuming the rotation along the angular momentum vector, for the rotation phases of Apophis
at the time of observation are uncertain due to its non-principal axis rotation.
Through the rotationally averaged procedure, the influence arising from the tumbling rotation
would somewhat decrease, but the thermal properties can be well revealed
from the seasonal effect, because we have observations at different epochs.

\section{Analysis and Results}
As mentioned above, we firstly assume the spectral emissivity to be constant 0.9
for each observed wavelength, and obtain the reduced $\chi^{2}_{\rm r}$ for each input
$\Gamma$ and $f_{\rm r}$ (showed in Table \ref{fit1}).
\begin{table}[htbp]\footnotesize
\renewcommand\arraystretch{1.1}
\centering
\caption{ATPM fitting results to the observations with constant spectral emissivity $=0.9$ for each wavelength.}
\label{fit1}
\begin{tabular}{@{}p{0.5cm}ccccccccccccccc@{}}
\hline
 Roughness &\multicolumn{12}{c}{Thermal inertia $\Gamma$ ($\rm~Jm^{-2}s^{-0.5}K^{-1}$)} \\
 fraction & \multicolumn{2}{c}{0} & \multicolumn{2}{c}{50}
          & \multicolumn{2}{c}{100} & \multicolumn{2}{c}{150}
          & \multicolumn{2}{c}{200} & \multicolumn{2}{c}{250}
          & \multicolumn{2}{c}{300}  \\
 $f_{\rm R}$ & $p_{\rm v}$ & $\chi^{2}_{\rm r}$ & $p_{\rm v}$ & $\chi^{2}_{\rm r}$
             & $p_{\rm v}$ & $\chi^{2}_{\rm r}$ & $p_{\rm v}$ & $\chi^{2}_{\rm r}$
             & $p_{\rm v}$ & $\chi^{2}_{\rm r}$ & $p_{\rm v}$ & $\chi^{2}_{\rm r}$
             & $p_{\rm v}$ & $\chi^{2}_{\rm r}$ \\
\hline
0.00 &0.316 & 4.64 &0.342 & 5.49 &0.346 & 5.94 &0.347 & 6.33 &0.347 & 6.70 &0.347 & 7.08 &0.347 & 7.42  \\
0.05 &0.313 & 4.46 &0.335 & 5.28 &0.338 & 5.73 &0.339 & 6.12 &0.339 & 6.49 &0.339 & 6.88 &0.339 & 7.23  \\
0.10 &0.312 & 4.29 &0.335 & 5.08 &0.337 & 5.53 &0.338 & 5.92 &0.338 & 6.29 &0.338 & 6.68 &0.338 & 7.04  \\
0.15 &0.312 & 4.12 &0.334 & 4.89 &0.337 & 5.33 &0.337 & 5.72 &0.337 & 6.10 &0.337 & 6.49 &0.337 & 6.86  \\
0.20 &0.311 & 3.97 &0.333 & 4.70 &0.336 & 5.13 &0.336 & 5.52 &0.336 & 5.90 &0.336 & 6.31 &0.336 & 6.67  \\
0.25 &0.310 & 3.84 &0.333 & 4.53 &0.335 & 4.95 &0.336 & 5.34 &0.335 & 5.72 &0.335 & 6.12 &0.335 & 6.49  \\
0.30 &0.310 & 3.71 &0.332 & 4.36 &0.334 & 4.77 &0.335 & 5.15 &0.334 & 5.53 &0.334 & 5.94 &0.334 & 6.32  \\
0.35 &0.309 & 3.60 &0.331 & 4.20 &0.333 & 4.59 &0.334 & 4.98 &0.333 & 5.36 &0.333 & 5.77 &0.333 & 6.15  \\
0.40 &0.308 & 3.49 &0.330 & 4.04 &0.332 & 4.43 &0.333 & 4.80 &0.332 & 5.18 &0.332 & 5.60 &0.332 & 5.98  \\
0.45 &0.307 & 3.40 &0.330 & 3.89 &0.331 & 4.27 &0.332 & 4.64 &0.331 & 5.01 &0.331 & 5.43 &0.331 & 5.82  \\
0.50 &0.306 & 3.32 &0.329 & 3.75 &0.330 & 4.11 &0.330 & 4.48 &0.330 & 4.85 &0.330 & 5.27 &0.329 & 5.65  \\
0.55 &0.305 & 3.25 &0.328 & 3.62 &0.329 & 3.96 &0.329 & 4.32 &0.329 & 4.69 &0.328 & 5.11 &0.328 & 5.50  \\
0.60 &0.304 & 3.19 &0.326 & 3.49 &0.328 & 3.82 &0.328 & 4.17 &0.327 & 4.54 &0.327 & 4.95 &0.326 & 5.34  \\
0.65 &0.303 & 3.14 &0.325 & 3.37 &0.327 & 3.68 &0.326 & 4.02 &0.326 & 4.39 &0.326 & 4.80 &0.325 & 5.19  \\
0.70 &0.302 & 3.10 &0.324 & 3.26 &0.325 & 3.55 &0.325 & 3.88 &0.324 & 4.24 &0.324 & 4.65 &0.323 & 5.04  \\
0.75 &0.300 & 3.07 &0.323 & 3.15 &0.324 & 3.43 &0.323 & 3.75 &0.323 & 4.10 &0.322 & 4.51 &0.322 & 4.90  \\
0.80 &0.299 & 3.05 &0.321 & 3.05 &0.322 & 3.31 &0.322 & 3.61 &0.321 & 3.96 &0.321 & 4.37 &0.320 & 4.76  \\
0.85 &0.297 & 3.05 &0.319 & 2.96 &0.320 & 3.19 &0.320 & 3.49 &0.319 & 3.83 &0.319 & 4.23 &0.318 & 4.62  \\
0.90 &0.296 & 3.05 &0.318 & 2.87 &0.319 & 3.08 &0.318 & 3.37 &0.317 & 3.70 &0.317 & 4.10 &0.316 & 4.48  \\
0.95 &0.294 & 3.06 &0.316 & 2.79 &0.317 & 2.98 &0.316 & 3.25 &0.315 & 3.57 &0.315 & 3.97 &0.314 & 4.35  \\
1.00 &0.290 & 3.08 &0.318 & 2.71 &0.319 & 2.88 &0.318 & 3.14 &0.317 & 3.45 &0.316 & 3.84 &0.315 & 4.22  \\
\hline
\end{tabular}
\end{table}

\begin{figure}[htbp]
\includegraphics[scale=0.71]{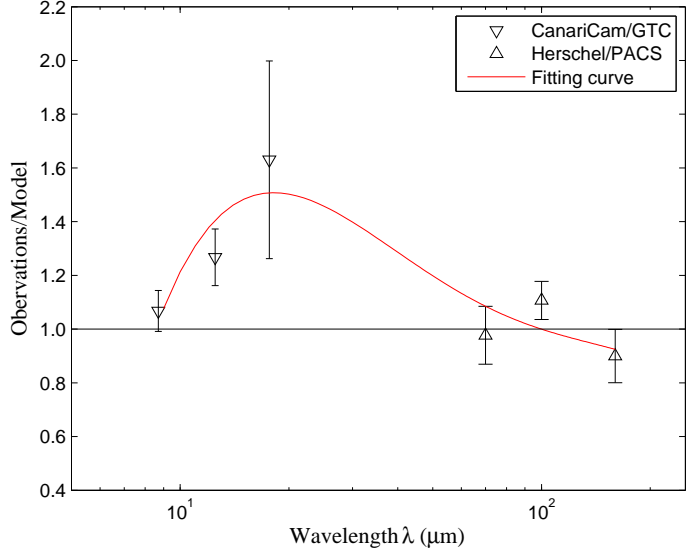}
\centering
\caption{The observation/model ratios as a function of wavelength
  for $\Gamma=50\rm~Jm^{-2}s^{-0.5}K^{-1}$, $f_{\rm r}=1.0$, $p_{\rm v}=0.318$,
  and $D_{\rm eff}=382\rm~m$ when using $\epsilon(\lambda)\equiv0.9$. The ratios with their errors
  at each wavelength are all error-weighted average results. The red fitting curve is obtained from a
  4th degree polynomial fitting for $x=\log10(\lambda)$, $y=F_{\rm obs}/F_{\rm model}$.
  }\label{rsp1}
\end{figure}

From Table \ref{fit1}, we can see that low thermal inertia and high roughness
fraction tend to fit better to the observations; the best-fit solution, corresponding
to the lowest reduced $\chi^{2}_{\rm r}$ arises at about $p_{\rm v}=0.318$
$\Gamma=50\rm~Jm^{-2}s^{-0.5}K^{-1}$ and $f_{\rm r}=1.0$. With this result, we
plot the ratio of 'observation/model' to examine how these theoretical model
results match the observations at various wavelengths (Figure \ref{rsp1}).

In Figure \ref{rsp1}, we can see that the ratios at each wavelength are not randomly
distributed around 1.0, but reveal a wavelength dependent feature like the red fitting
curve, where the maximum ratio may arise around $20~\mu m$. This feature implies that
the spectral emissivity used in Equation (\ref{fmodel}) should not be a constant
$\epsilon(\lambda)\equiv0.9$, but should be wavelength dependent, where the maximum
emissivity may arise around $20~\mu m$. Thus we should introduce a wavelength dependent
spectral emissivity $\epsilon(\lambda)$ to do the fitting process again.

\citet{Muller1998} reported the wavelength dependent spectral emissivity of asteroid
Ceres and Vesta. As the spectral feature of Vesta may be much closer to Apophis than
Ceres, so we imagine that the Apophis' emissivity changes with wavelength in a similar
way like that of asteroid Vesta \citep{Muller1998} as showed in Table \ref{ems}. With
these wavelength dependent emissivity, we did get better solutions.
\begin{table}[htbp]
 \centering
 \renewcommand\arraystretch{1.5}
 \caption{Assumed Emissivity for different wavelength.}
 \label{ems}
 \begin{tabular}{@{}lrc@{}}
 \hline
 & Wavelength & Emissivity \\
 \hline
 & 8.7, 12.5  $\mu m$ & 0.90 \\
 & 17.65 $\mu m$      & 0.96 \\
 & 70, 100  $\mu m$   & 0.80 \\
 & 160 $\mu m$        & 0.75 \\
 \hline
\end{tabular}
\end{table}

\begin{figure}[htbp]
\includegraphics[scale=0.75]{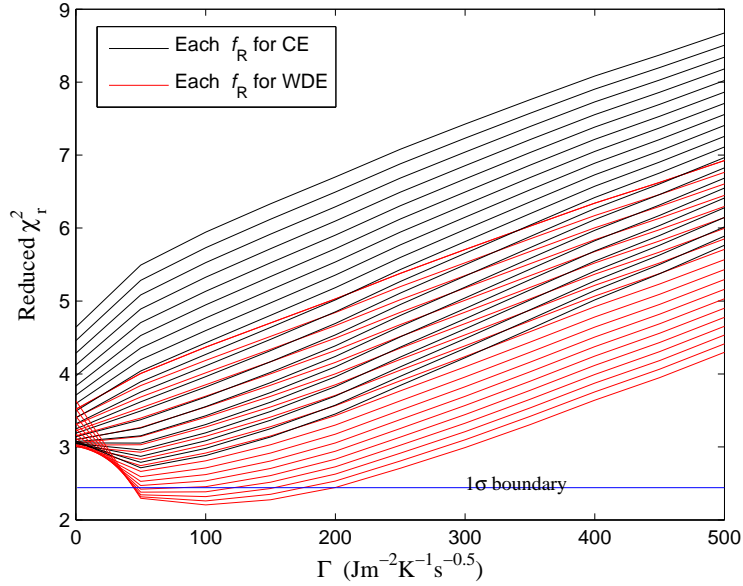}
  \centering
  \caption{$\Gamma\sim\chi^2_{\rm r}$ curves for both the CE (constant emissivity )
  and WDE (wavelength dependent emissivity) cases considering $f_{\rm r}=0.0\sim1.0$.
  }\label{vem}
\end{figure}

Table \ref{fit2} summarises the reduced $\chi^2_{\rm r}$ obtained from each input
pair of thermal inertia and roughness fraction when using the emissivity given in
Table \ref{ems}. We can see that, in this case, high roughness fraction tends to
fit better as well. But the best-fit thermal inertia and geometric albedo shift to
$\Gamma=100\rm~Jm^{-2}s^{-0.5}K^{-1}$ and $p_{\rm v}=0.286$ respectively. We compared
the reduced $\chi^2_{\rm r}$ obtained from the two case CE and WDE in Figure \ref{vem},
where the black curves represent the reduced $\chi^{2}$ obtained from constant
emissivity input while the red curves stand for wavelength dependent emissivity.
We can see that the obtained minimum reduced $\chi^{2}_{\rm r}$ can be significantly
lower when the WDE are used to fit observations, indicating that the spectral emissivity
of Apophis should be wavelength dependent.

\begin{landscape}
\begin{table}[htbp]\footnotesize
\vskip 20pt
\renewcommand\arraystretch{1.0}
\centering
\caption{ATPM fitting results to the observations with wavelength dependent emissivity.}
\label{fit2}
\begin{tabular}{@{}p{0.5cm}cccccccccccccccccccccc@{}}
\hline
 Roughness &\multicolumn{22}{c}{Thermal inertia $\Gamma$ ($\rm~Jm^{-2}s^{-0.5}K^{-1}$)} \\
 fraction & \multicolumn{2}{c}{0} & \multicolumn{2}{c}{50}
          & \multicolumn{2}{c}{100} & \multicolumn{2}{c}{150}
          & \multicolumn{2}{c}{200} & \multicolumn{2}{c}{250}
          & \multicolumn{2}{c}{300} & \multicolumn{2}{c}{350}
          & \multicolumn{2}{c}{400} & \multicolumn{2}{c}{450}
          & \multicolumn{2}{c}{500} \\
 $f_{\rm R}$ & $p_{\rm v}$ & $\chi^{2}_{\rm r}$ & $p_{\rm v}$ & $\chi^{2}_{\rm r}$
             & $p_{\rm v}$ & $\chi^{2}_{\rm r}$ & $p_{\rm v}$ & $\chi^{2}_{\rm r}$
             & $p_{\rm v}$ & $\chi^{2}_{\rm r}$ & $p_{\rm v}$ & $\chi^{2}_{\rm r}$
             & $p_{\rm v}$ & $\chi^{2}_{\rm r}$ & $p_{\rm v}$ & $\chi^{2}_{\rm r}$
             & $p_{\rm v}$ & $\chi^{2}_{\rm r}$ & $p_{\rm v}$ & $\chi^{2}_{\rm r}$
             & $p_{\rm v}$ & $\chi^{2}_{\rm r}$ \\
\hline
0.00 &0.283 & 3.64 &0.305 & 4.18 &0.308 & 4.55 &0.309 & 4.89 &0.309 & 5.22 &0.309 & 5.57 &0.308 & 5.89 &0.308 & 6.21 &0.307 & 6.52 &0.306 & 6.80 &0.306 & 7.09 \\
0.05 &0.284 & 3.52 &0.304 & 4.01 &0.306 & 4.37 &0.307 & 4.70 &0.307 & 5.03 &0.307 & 5.38 &0.306 & 5.70 &0.306 & 6.03 &0.306 & 6.34 &0.305 & 6.63 &0.305 & 6.93 \\
0.10 &0.284 & 3.40 &0.304 & 3.84 &0.306 & 4.19 &0.307 & 4.52 &0.306 & 4.84 &0.306 & 5.20 &0.306 & 5.52 &0.305 & 5.85 &0.305 & 6.17 &0.304 & 6.46 &0.304 & 6.76 \\
0.15 &0.284 & 3.31 &0.304 & 3.68 &0.306 & 4.02 &0.306 & 4.34 &0.306 & 4.66 &0.306 & 5.02 &0.305 & 5.35 &0.305 & 5.68 &0.304 & 6.00 &0.304 & 6.30 &0.303 & 6.60 \\
0.20 &0.283 & 3.22 &0.303 & 3.53 &0.305 & 3.85 &0.305 & 4.17 &0.305 & 4.49 &0.305 & 4.84 &0.305 & 5.18 &0.304 & 5.51 &0.304 & 5.84 &0.303 & 6.14 &0.303 & 6.45 \\
0.25 &0.283 & 3.15 &0.303 & 3.40 &0.305 & 3.70 &0.305 & 4.00 &0.305 & 4.32 &0.304 & 4.68 &0.304 & 5.01 &0.303 & 5.35 &0.303 & 5.68 &0.302 & 5.98 &0.302 & 6.30 \\
0.30 &0.282 & 3.10 &0.302 & 3.27 &0.304 & 3.55 &0.304 & 3.84 &0.304 & 4.16 &0.304 & 4.51 &0.303 & 4.85 &0.303 & 5.19 &0.302 & 5.52 &0.301 & 5.82 &0.301 & 6.14 \\
0.35 &0.282 & 3.06 &0.302 & 3.14 &0.303 & 3.41 &0.304 & 3.69 &0.303 & 4.00 &0.303 & 4.35 &0.302 & 4.69 &0.302 & 5.03 &0.301 & 5.37 &0.301 & 5.67 &0.300 & 6.00 \\
0.40 &0.281 & 3.03 &0.301 & 3.03 &0.303 & 3.27 &0.303 & 3.55 &0.302 & 3.85 &0.302 & 4.20 &0.301 & 4.54 &0.301 & 4.88 &0.300 & 5.22 &0.300 & 5.52 &0.299 & 5.85 \\
0.45 &0.281 & 3.01 &0.301 & 2.93 &0.302 & 3.15 &0.302 & 3.41 &0.301 & 3.71 &0.301 & 4.05 &0.300 & 4.39 &0.300 & 4.73 &0.299 & 5.07 &0.299 & 5.38 &0.298 & 5.71 \\
0.50 &0.280 & 3.01 &0.300 & 2.83 &0.301 & 3.03 &0.301 & 3.28 &0.300 & 3.57 &0.300 & 3.91 &0.299 & 4.24 &0.299 & 4.58 &0.298 & 4.92 &0.298 & 5.23 &0.297 & 5.57 \\
0.55 &0.279 & 3.02 &0.299 & 2.74 &0.300 & 2.92 &0.300 & 3.15 &0.299 & 3.43 &0.299 & 3.77 &0.298 & 4.10 &0.298 & 4.44 &0.297 & 4.78 &0.296 & 5.09 &0.296 & 5.43 \\
0.60 &0.279 & 3.04 &0.298 & 2.66 &0.299 & 2.81 &0.299 & 3.03 &0.298 & 3.30 &0.298 & 3.63 &0.297 & 3.96 &0.297 & 4.30 &0.296 & 4.64 &0.295 & 4.95 &0.295 & 5.29 \\
0.65 &0.278 & 3.07 &0.298 & 2.59 &0.298 & 2.71 &0.298 & 2.92 &0.297 & 3.18 &0.297 & 3.50 &0.296 & 3.82 &0.295 & 4.16 &0.295 & 4.51 &0.294 & 4.82 &0.294 & 5.16 \\
0.70 &0.277 & 3.12 &0.297 & 2.53 &0.297 & 2.62 &0.297 & 2.81 &0.296 & 3.06 &0.296 & 3.38 &0.295 & 3.69 &0.294 & 4.03 &0.294 & 4.37 &0.293 & 4.69 &0.292 & 5.03 \\
0.75 &0.276 & 3.17 &0.296 & 2.47 &0.296 & 2.54 &0.296 & 2.71 &0.295 & 2.94 &0.294 & 3.25 &0.293 & 3.57 &0.293 & 3.90 &0.292 & 4.24 &0.291 & 4.56 &0.291 & 4.90 \\
0.80 &0.275 & 3.24 &0.294 & 2.42 &0.295 & 2.46 &0.294 & 2.61 &0.293 & 2.83 &0.293 & 3.14 &0.292 & 3.44 &0.291 & 3.78 &0.291 & 4.12 &0.290 & 4.43 &0.290 & 4.78 \\
0.85 &0.274 & 3.32 &0.293 & 2.38 &0.294 & 2.38 &0.293 & 2.52 &0.292 & 2.73 &0.291 & 3.02 &0.290 & 3.33 &0.290 & 3.65 &0.289 & 3.99 &0.288 & 4.31 &0.288 & 4.65 \\
0.90 &0.273 & 3.42 &0.292 & 2.34 &0.292 & 2.32 &0.291 & 2.43 &0.290 & 2.63 &0.290 & 2.91 &0.289 & 3.21 &0.288 & 3.53 &0.288 & 3.87 &0.287 & 4.18 &0.286 & 4.53 \\
0.95 &0.272 & 3.52 &0.291 & 2.32 &0.291 & 2.26 &0.290 & 2.35 &0.289 & 2.53 &0.288 & 2.81 &0.287 & 3.10 &0.286 & 3.42 &0.286 & 3.76 &0.285 & 4.07 &0.285 & 4.41 \\
1.00 &0.263 & 3.63 &0.286 & 2.30 &0.286 & 2.21 &0.285 & 2.28 &0.284 & 2.44 &0.283 & 2.71 &0.282 & 2.99 &0.281 & 3.31 &0.280 & 3.64 &0.279 & 3.95 &0.278 & 4.30 \\
\hline
\end{tabular}
\end{table}
\end{landscape}

Since we can not obtain best solutions by assuming constant emissivity, then we
have no reason to fit the observations with ATPM by using constant emissivity.
Therefore we adopt the wavelength dependent emissivity similar to that of asteroid
Vesta to further derive the surface thermal inertia, roughness fraction and geometric
albedo here.

\begin{figure}[htbp]
\includegraphics[scale=0.75]{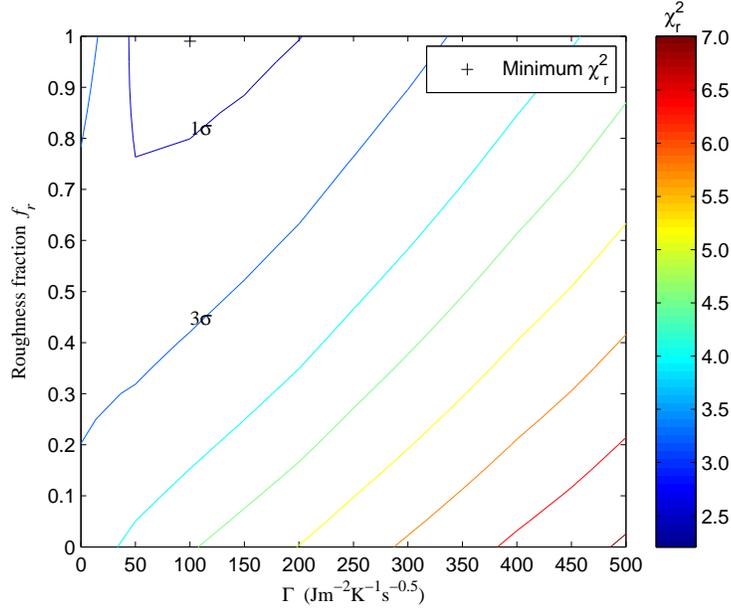}
  \centering
  \caption{$\chi^{2}_{\rm r}$ ($f_{\rm r}$, $\Gamma$) contour according to Table \ref{fit2}.
  The color (from blue to red) means the increase of profile of $\chi^{2}_{\rm r}$.
  1$\sigma$ corresponds to $\Delta\chi^2_{\rm r}=0.252$, while 3$\sigma$
  corresponds to $\Delta\chi^2_{\rm r}=1.014$ \citep{Press2007}.
  }\label{contour}
\end{figure}

Figure \ref{contour} shows the contour of $\chi^{2}_{\rm r}$ ($f_{\rm r}$, $\Gamma$)
according to Table \ref{fit2}, where the color variation from blue to red means
the increase of the reduced $\chi^{2}_{\rm r}$. The deep blue curve labelled by
1$\sigma$ corresponds to $\Delta\chi^2_{\rm r}=0.252$ from the
minimum $\chi^{2}_{\rm r}$, constraining the scale of the free parameters with
possibility about $68.3\%$, whereas the blue curve labelled by 3$\sigma$ refers to
$\Delta\chi^2_{\rm r}=1.014$, giving the scale of free parameters with
possibility about $99.73\%$ \citep{Press2007}.

\begin{figure}[htbp]
\includegraphics[scale=0.75]{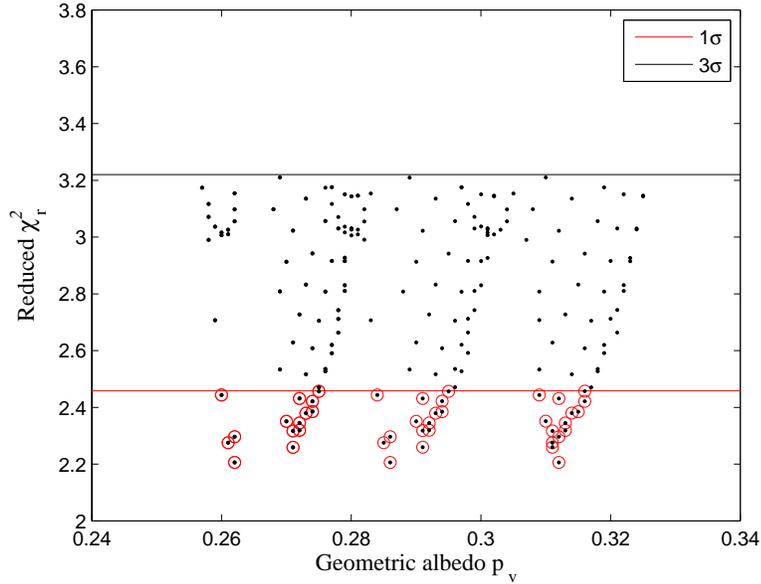}
  \centering
  \caption{$p_{\rm v}\sim\chi^2_{\rm reduced}$ profiles fit to the observations
  in consideration of the derived 1$\sigma$ and 3$\sigma$ range of $f_{\rm r}$ and $\Gamma$
  and the absolute visual magnitude $H_{\rm v}=19.09\pm0.19$.
  }\label{galbedo}
\end{figure}

Figure \ref{galbedo} shows the $p_{\rm v}\sim\chi^2_{\rm r}$ obtained in consideration
of the above derived $\Gamma$ and $f_{\rm r}$ as well as the absolute visual magnitude
$H_{\rm v}=19.09\pm0.19$, where the 1$\sigma$ and 3$\sigma$ limits are the same as above.
The scale of absolute visual magnitude $H_{\rm v}$ does not affect the distribution of
reduced $\chi^2_{\rm r}$ derived from each pair of thermal inertia and roughness fraction,
but has significant influence on the corresponding geometric albedo and effective diameter.
Thus we do not treat $H_{\rm v}$ as free parameters, but consider its influence by only
using the upper limit, mid-value and lower limit of $H_{\rm v}$, each was adopted to be a
constant in the fitting process. And the final results of geometric albedo and effective
diameter are constrained by considering three cases of $H_{\rm v}$ together, leading to
the 1$\sigma$ and 3$\sigma$ scales non-Gaussian.

Using the assumed wavelength dependent emissivity in Table \ref{ems},
we can derive thermal inertia $\Gamma$, roughness fraction $f_{\rm r}$, geometric albedo $p_{\rm v}$ and
effective diameter $D_{\rm eff}$,  where are considered be free parameters in the fitting procedure,
on the likely 1$\sigma$ and 3$\sigma$ scale. And we summarize the relevant results
in Table \ref{results}.

\begin{table}[htbp]
 \centering
 \renewcommand\arraystretch{2}
 \caption{The derived parameters for Apophis.}
 \label{results}
 \begin{tabular}{@{}lcc@{}}
 \hline
  Properties & 1$\sigma$ scale & 3$\sigma$ scale \\
 \hline
  Thermal inertia $\Gamma$($\rm~Jm^{-2}s^{-0.5}K^{-1}$) & $100^{+100}_{-52}$ & $100^{+240}_{-100}$\\
  Roughness fraction $f_{\rm r}$   & 0.78$\sim$1.0 & 0.2$\sim$1.0 \\
  Geometric albedo $p_{\rm v}$   & $0.286^{+0.03}_{-0.026}$ & $0.286^{+0.039}_{-0.029}$ \\
  Effective diameter $D_{\rm eff}$ (m) & $378^{+19}_{-25}$ & $378^{+27}_{-29}$\\
 \hline
\end{tabular}
\end{table}

\section{Discussion and Conclusion}
In this work, the mid-infrared and far-infrared data of Apophis observed by
CanariCam on Gran Telescopio CANARIAS and PACS of Herschel are analysed by
the Advanced thermophysical model (ATPM), where four parameters, including
thermal emissivity, thermal inertia, roughness fraction and geometric albedo,
are investigated. We found that the thermal emissivity of Apophis should be
wavelength dependent, and using a similar emissivity like that of Vesta could
obtain better degree of fitting. As a result, we derive the thermal inertia,
roughness fraction, geometric albedo and effective diameter of Apophis in a
possible 1$\sigma$ scale of
$\Gamma=100^{+100}_{-52}\rm~Jm^{-2}s^{-0.5}K^{-1}$, $f_{\rm r}=0.78\sim1.0$,
$p_{\rm v}=0.286^{+0.030}_{-0.026}$, $D_{\rm eff}=378^{+19}_{-25}\rm~m$,
and 3$\sigma$ scale of $\Gamma=100^{+240}_{-100}\rm~Jm^{-2}s^{-0.5}K^{-1}$,
$f_{\rm r}=0.2\sim1.0$, $p_{\rm v}=0.286^{+0.039}_{-0.029}$,
$D_{\rm eff}=378^{+27}_{-29}\rm~m$. The derived thermal inertia
supports the result of \citet{Licandro2016} despite a little lower. The best
fit high roughness is different from the result of \citet{Licandro2016},
but is inconsistent with with the work of \citet{Muller2014} where the best
fit solution also supports a high roughness.

\begin{figure}[htbp]
\includegraphics[scale=0.75]{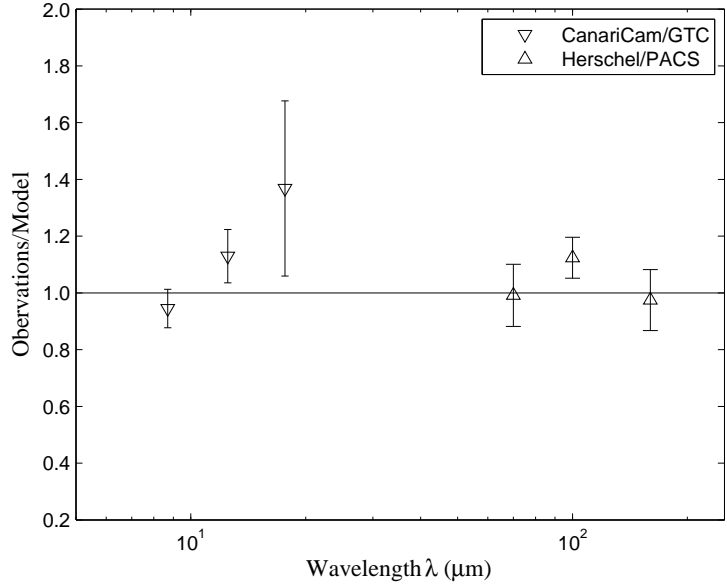}
\centering
\caption{The observation/model ratios as a function of wavelength for
  $\Gamma=100\rm~Jm^{-2}s^{-0.5}K^{-1}$, $f_{\rm r}=1.0$, $p_{\rm v}=0.286$
  and $D_{\rm eff}=378\rm~m$ in consideration of wavelength dependent emissivity.
  The ratios with their errors at each wavelength are all error-weighted average results.
  }\label{rsp2}
\end{figure}

To verify the the reliability of our fitting procedure and derived outcomes, we
employ the ratio of 'observation/model' \citep{Muller2005,Muller2011,Muller2012}
to examine how these theoretical model results match the observations at various
wavelengths (see Figure \ref{rsp2}). In Figure \ref{rsp2}, the observation/Model
ratios are shown at each observational wavelength for $f_{\rm r}=1.0$,
$\Gamma=100\rm~Jm^{-2}s^{-0.5}K^{-1}$, $p_{\rm v}=0.286$ and $D_{\rm eff}=378\rm~m$.
The ratios are distributed more nearly around 1.0 compared to Figure \ref{rsp1},
despite the ratios at $17.65\rm~\mu m$ move relatively larger than unity which again
support the idea that Apophis' spectral emissivity should be wavelength dependent.
However, we can not know how exactly the spectral emissivity varies with wavelength
at present, for the available spectral data is not enough. If, in the future, more
spectral data observed from mid-infrared to far-infrared are obtained, we may find
out how the emissivity of Apophis depends on wavelength.

Given the low gravitational acceleration on the surface of asteroids, it was
generally thought that regolith formation would not be possible, especially
on small-size asteroids. The statistical results of thermal inertia versus the size
of asteroids in the work of \citet{Delbo2007} suggest that small-size asteroid
should have high thermal inertia, indicating rocky surface, also support the
above idea. But regolith on the small-size asteroid Itokawa was detected by the
Hayabusa spacecraft, which suggests that regolith formation on asteroids, even
small-size asteroids is also possible. The low thermal inertia result of Apophis
in this work somewhat adds a good support to the idea that regolith formation may
be not dominated by the size of asteroid. If a small asteroid has experienced enough
long process of space weathering, the presence of regolith should be also reasonable.
On the other hand, the size of Apophis is similar to asteroid Itokawa indicative of
similar dynamical lifetime in principle. But the derived thermal inertia of Apophis
is much lower than the observed thermal inertia of Itokawa \citep{Muller2005},
indicating Apophis may actually either exit a longer lifetime or has less regolith
migration process on the surface, making more dust produced and retained over the
surface. We notice that Apophis has much slower rotation $P_{\rm rot}\approx30.56$ hr
than Itokawa ($P_{\rm rot}\approx12.13$ hr), which may account for Apophis' lower
thermal inertia, for the produced dust should be more easy to accumulate on a
slow-rotating asteroid.

Generally rough surface would generate stronger Yarkovsky orbit drift than
smooth surface \citep{Rozitis2012}. Thus the Yarkovsky effect induced decreasing
rate of Apophis' semimajor axis should be even larger than the predicted value
of \citet{Vokrouhlicky2015}, because our results show Apophis tends to have a
high-roughness surface, which should be taken into account to predict its orbit
movement. Besides it is possible that Apophis may be one Vesta's fragment delivered
from the Main Belt by Yarkovsky effect, in consideration of its long existence time
suggested by low thermal inertia, and similar surface properties to Vesta implied
by albedo and thermal emissivity.

In conclusion, when we attempt to investigate surface thermophysical properties of
a target asteroid by utilizing thermal infrared radiometric method to fit observed
data from mid-infrared to far-infrared, constant emissivity may be not always the
best choice. For asteroid Apophis, the combined data of mid-infrared of  CanariCam
and far-infrared of Herschel reveals possible wavelength dependent thermal emissivity
from $8.70~\mu m$ to $160~\mu m$, where the maximum emissivity may arise around
$20~\mu m$ like that of Vesta, because this kind of emissivity can achieve better
degree of fitting through the use of Advanced thermophysical model. Besides the
derived results of low thermal inertia $\Gamma=100^{+240}_{-100}\rm~Jm^{-2}s^{-0.5}K^{-1}$
and high roughness fraction $f_{\rm r}=0.2\sim1.0$ indicate that Apophis may have
experienced a long process of space weathering, but less regolith migration process,
making the produced dust able to stay over its surface. These new deductions would be
important for predicting Apophis' orbital movement and potential Earth impact probability
so as to establish artificial mechanism to avoid the probable impact.

\begin{acknowledgements}
We thank the anonymous referee for good comments and suggestions
that helped to improve the manuscript. This work is financially supported by
National Natural Science Foundation of China (Grants No. 11473073, 11403105, 11661161013),
the Science and Technology Development Fund of Macau (Grants No. 039/2013/A2, 017/2014/A1),
the innovative and interdisciplinary program by CAS (Grant No. KJZD-EW-Z001),
the Natural Science Foundation of Jiangsu Province
(Grant No. BK20141509), and the Foundation of Minor Planets of Purple Mountain
Observatory.
\end{acknowledgements}

\bibliographystyle{named}

\begin{thebibliography}{00}

\bibitem[\protect\citeauthoryear{Abe et al.}{2006}]{Abe2006}
Abe, S., Mukai, T., Hirata, N., et al. 2006, Science, 312, 1344

\bibitem[\protect\citeauthoryear{Binzel et al.}{2009}]{Binzel2009}
Binzel, R. P., Rivkin, A. S., Thomas, C. A., et al. 2009, Icarus, 200, 480

\bibitem[\protect\citeauthoryear{Britt et al.}{2002}]{Britt2002}
Britt, D. T., Yeomans, D., Housen, K., \& Consolmagno, G. 2002, Asteroid Density, Porosity, and Structure.
In \emph{Asteroids III}, eds. W. F. Bottke Jr., A. Cellino, P. Paolicchi, \& R. P. Binzel (Tucson:
University of Arizona Press), 485

\bibitem[\protect\citeauthoryear{Britt \& Consolmagno}{2003}]{Britt2003}
Britt, D. T., \& Consolmagno, G. 2003, Meteoritics, 38, 1161

\bibitem[\protect\citeauthoryear{Bottke et al.}{2006}]{Bottke2006}
Bottke, W.F., Vokrouhlick$\rm \acute{y}$, D., Rubincam, D.P., Nesvorn$\rm \acute{y}$, D., 2006. The Yarkovsky
and YORP effects: Implications for asteroid dynamics. Annu. Rev. Earth Planet. Sci. 34, 157-191

\bibitem[\protect\citeauthoryear{Bottke et al.}{2015}]{Bottke2015}
Bottke, W.F., et al. 2015, Icarus, 247, 191-217

\bibitem[\protect\citeauthoryear{Bowell et al.}{1989}]{Bowell}
Bowell, E., Hapke, B., Domingue, D., et al., 1989,  Application of photometric models to asteroids.
In \emph{Asteroids II}, pp. 524-556

\bibitem[\protect\citeauthoryear{Chamberlain et al.}{2007}]{Chamberlain2007}
Chamberlain, M.A., Lovell, A.J., Sykes, M.V., 2007, Icarus, 192, 448-459

\bibitem[\protect\citeauthoryear{Chesley}{2005}]{Chesley2005}
Chesley, S.R., 2005. Potential impact detection for near-Earth asteroids: The case of
99942 Apophis (2004 MN 4). In: IAU Symposium No. 229, pp. 215-228.

\bibitem[\protect\citeauthoryear{Conel}{1969}]{Conel1969}
Conel, James E., 1969, Experimental results and a cloudy atmospheric model of spectral emission from condensed
particulate mediums, JGR, 74, 1614-1634

\bibitem[\protect\citeauthoryear{Delbo et al.}{2007}]{Delbo2007}
Delbo, M., Oro, A., Harris, A.W., Mottola, S., \& Muller, M., 2007, Icarus, 190, 236-249

\bibitem[\protect\citeauthoryear{Farnocchia et al.}{2013}]{Farnocchia2013}
Farnocchia, D. et al., 2013. Yarkovsky-driven impact risk analysis for Asteroid
(99942) Apophis. Icarus 224, 192-200

\bibitem[\protect\citeauthoryear{Fujiwara et al.}{2006}]{Fujiwara2006}
Fujiwara, A., Kawaguchi, J., Yeomans, D.K., et al, 2006, Science, 312, 1330-1334

\bibitem[\protect\citeauthoryear{Fowler \& Chillemi}{1992}]{Fowler1992}
Fowler, J.W., \& Chillemi, J.R., 1992, IRAS asteroids data processing. In The IRAS Minor Planet Survey, pp. 17-43

\bibitem[\protect\citeauthoryear{Gilmore et al.}{2004}]{Gilmore2004}
Gilmore, A.C. et al., 2004. 2004 MN4. Minor Planet Electronic Circulars Y, 25

\bibitem[\protect\citeauthoryear{Gilmore et al.}{2008}]{Gilmore2008}
Giorgini, J.D., Benner, L.A.M., Ostro, S.J., Nolan, M.C., Busch, M.W., 2008. Predicting
the Earth encounters of (99942)Apophis. Icarus 193, 1-19

\bibitem[\protect\citeauthoryear{Lagerros}{1996 I}]{Lagerros1996I}
Lagerros, J.S.V., 1996, A\&A, 310, 1011-1020

\bibitem[\protect\citeauthoryear{Lagerros}{1996 II}]{Lagerros1996II}
Lagerros, J.S.V., 1996, A\&A, 315, 625-632

\bibitem[\protect\citeauthoryear{Licandro et al.}{2016}]{Licandro2016}
Licandro, J., M\"{u}ller, T.G., Alvarez, C. et al, 2016, A\&A, 585, A10

\bibitem[\protect\citeauthoryear{M\"{u}ller \& Lagerros}{1998}]{Muller1998}
M\"{u}ller, T.G., \& Lagerros, J.S.V., 1998, A\&A, 338, 340-352

\bibitem[\protect\citeauthoryear{M\"{u}ller}{2002}]{Muller2002}
M\"{u}ller, T.G., 2002, Meteor. Planet. Sci., 37, 1919

\bibitem[\protect\citeauthoryear{M\"{u}ller et al.}{2005}]{Muller2005}
M\"{u}ller, T.G., Sekiguchi, T., Kaasalainen, M., Abe, M., \& Hasegawa, S., 2005, A\&A, 443, 347

\bibitem[\protect\citeauthoryear{M\"{u}ller et al.}{2011}]{Muller2011}
M\"{u}ller, T.G., Durech, J., Hasegawa, S., et al., 2011, A\&A, 525, A145

\bibitem[\protect\citeauthoryear{M\"{u}ller et al.}{2012}]{Muller2012}
M\"{u}ller, T.G., O'Rourke, L., Barucci, A.M., et al., 2012, A\&A, 548, A36

\bibitem[\protect\citeauthoryear{M\"{u}ller et al.}{2014}]{Muller2014}
M\"{u}ller, T.G., Kiss, C., Scheirich, P., et al., 2014, A\&A, 566, A22

\bibitem[\protect\citeauthoryear{Opeil et al.}{2012}]{Opeil2012}
Opeil, C.P., Consolmagno, G.J., Safarik, D.J., Britt, D.T., 2012, Meteorit. Planet. Sci. 47, 319-329

\bibitem[\protect\citeauthoryear{Ostro et al.}{1999}]{Ostro1999}
Ostro, S.J., Hudson, R.S., Rosema, K.D., et al., 1999, Icarus 137, 122-139

\bibitem[\protect\citeauthoryear{Pravec et al.}{2014}]{Pravec2014}
Pravec, P., Scheirich, P., \v{D}urech, J., et al., 2014, Icarus, 233, 48-60

\bibitem[\protect\citeauthoryear{Press et al.}{2007}]{Press2007}
Press W.H., et al., Numerical Recipes, 3th Edition, 2007, Cambridge University Press, NewYork, p. 815

\bibitem[\protect\citeauthoryear{Rozitis \& Green}{2011}]{Rozitis2011}
Rozitis, B., \& Green, S.F., 2011, MNRAS, 415, 2042

\bibitem[\protect\citeauthoryear{Rozitis \& Green}{2012}]{Rozitis2012}
Rozitis, B., \& Green, S.F., 2012, MNRAS, 423, 367

\bibitem[\protect\citeauthoryear{Saito et al.}{2006}]{Saito2006}
Saito J., Miyamoto H., Nakamura R. Detailed Images of Asteroid 25143
Itokawa from Hayabusa. Science, 2006, 312: 1341

\bibitem[\protect\citeauthoryear{Thuillot et al.}{2015}]{Thuillot2015}
Thuillot, W., Bancelin, D, Ivantsov, A., et al., 2015, A\&A, 583, A59

\bibitem[\protect\citeauthoryear{Vokrouhlick\'{y} et al.}{2015}]{Vokrouhlicky2015}
Vokrouhlick\'{y}, D., Farnocchia, D., \v{C}apek, D., et al., 2015, Icarus, 252, 277-283

\bibitem[\protect\citeauthoryear{Wang et al.}{2015}]{Wang2015}
Wang, N., Peng, Q.Y., Zhang, X.L., et al., 2015, MNRAS, 454, 3805-3809

\bibitem[\protect\citeauthoryear{Yu, Ji \& Wang}{2014}]{Yu2014}
Yu, L.L., Ji, J.H., \& Wang, S., 2014, MNRAS, 439, 3357-3370

\bibitem[\protect\citeauthoryear{Yu \& Ji}{2015}]{Yu2015}
Yu, L.L., \& Ji, J.H., 2015, MNRAS, 452, 368-375

\end{thebibliography}

\label{lastpage}

\end{document}